# Title: Nanoscale Imaging of Phonons and Reconfiguration in Topologically-Engineered, Self-Assembled Nanoparticle Lattices

**Authors:** Chang Qian[1]†, Ethan Stanifer[2]†, Zhan Ma[3], Binbin Luo[1], Chang Liu[1], Lehan Yao[1], Wenxiao Pan[3], Xiaoming Mao[2]*, Qian Chen[1,4,5,6]*

**Affiliations:**

[1]Department of Materials Science and Engineering, University of Illinois, Urbana, IL 61801, United States

[2]Department of Physics, University of Michigan, Ann Arbor, MI 48109, United States

[3]Department of Mechanical Engineering, University of Wisconsin-Madison, Madison, WI 53706, United States

[4]Department of Chemistry, University of Illinois, Urbana, IL 61801, United States

[5]Beckman Institute for Advanced Science and Technology, University of Illinois, Urbana, IL 61801, United States

[6]Materials Research Laboratory, University of Illinois, Urbana, IL 61801, United States

†These authors contribute equally to the work.

*Correspondence should be addressed to: maox@umich.edu; qchen20@illinois.edu

**Abstract:**

Topologically-engineered mechanical frames are important model constructs for architecture, machine mechanisms, and metamaterials. Despite significant advances in macroscopically fashioned frames, realization and phonon imaging of nanoframes have remained challenging. Here we extend for the first time the principles of topologically-engineered mechanical frames to lattices self-assembled from nanoparticles. Liquid-phase transmission electron microscopy images the vibrations of nanoparticles in self-assembled Maxwell and hexagonal lattices at the nanometer resolution, measuring a series of otherwise inaccessible properties such as phonon spectra and nonlinear lattice deformation paths. These properties are experimentally modulated by ionic strength, captured by our discrete mechanical model considering the complexity of nanoscale interactions and thermal fluctuations. The experiment–theory integration bridges mechanical metamaterials and colloidal self-assembly, opening new opportunities to manufacture phononic devices with solution processability, transformability, light weight, and emergent functions, at underexplored length, frequency, and energy scales.

**Main text:**

Topological design principles are employed in many systems to produce exotic properties, from the MAXXI museum featuring curved ramps and voids for shadow effects (*1*), shape-morphing actuators for soft robotics (*2*), to topological metamaterials (*3*) inspired by the discovery of topological insulators in condensed matter (*4*). In the last case, topology control can engineer phonons and photons with remarkable phenomena, from spin-momentum locking, unidirectional edge states, to reconfigurable waveguiding (*5*). For example, Maxwell lattices as a special class of topologically-engineered mechanical frames exhibit marginal stability, where degrees of freedom balance with constraints in the bulk (*6*). In contrast to close packing, Maxwell lattices are porous systems with structural degeneracy that is sub-extensive in system size, leading to transformability and topologically protected low-energy modes (i.e., floppy modes) (*3, 7*). These features are important for applications in shock absorption, stress focusing, and nonreciprocal wave transmission (*8, 9*). Due to such broad relevance, topologically-engineered mechanical frames have been extensively realized at the macroscopic scale (*7, 9, 10*) and more recently at the nanoscale (*11-13*) by top-down lithography or printing.

Meanwhile, self-assembly of colloidal nanoparticles (NPs) has attracted extensive efforts over the last few decades for their unique advantages to generate scalable, solution processable, and reconfigurable structures of exquisite topology control, thanks to the high tunability of NP shape, composition, and inter-NP interaction (*14, 15*). However, studying NP-assembled lattices as mechanical frames has remained unexplored, let alone mapping and understanding their phonon behaviors. On one hand, NP-assembled mechanical frames can cover unique frequency ranges (a few MHz to hundreds of GHz) and energy scales that are highly desirable to access in phonon manipulation for applications in optomechanical devices and thermal transport (note S1, table S1) (*16, 17*). NPs also promise unique potentials to make novel metamaterials by coupling their shape- and size-defined optical, electronic, catalytic, and magnetic properties with the lattices' topologically-encoded mechanical features (*14, 18, 19*). On the other hand, while digital camera, laser vibrometry, and first-principle models have been developed and used for macroscopic mechanical frames (*20, 21*), these methods do not apply to NP-assembled lattices due to insufficient imaging resolution or limited coverage on the frequency and energy range. Moreover, new theoretical advancements over existing framework of mechanical metamaterials are needed to assess and incorporate factors such as NP shape anisotropy, thermal fluctuation, as well as the directionality, range, anharmonicity, and multi-body nature of inter-NP interactions, which are emergent design parameters for NP-assembled mechanical frames.

Here we present the first experimental realization and theoretical framework of a self-assembled, topologically-engineered Maxwell lattice from NPs suspended in solution. The phonon behaviors and the consequent reconfiguration of the lattice upon thermal perturbation are mapped at the previously inaccessible nanometer resolution, enabled by adapting liquid-phase transmission electron microscopy (TEM) (*22-24*)—a recent breakthrough that images the real-space and real-time dynamics in solution under TEM—to phonon mode nanoscopy (PMN). For Maxwell lattices, one essential structural ingredient is a perfect hinge. The hinge constrains longitudinal motion while allowing freedom for rotation and structural degeneracy (Fig. 1a), which we realize experimentally at the nanoscale by employing interactions of anisotropic NPs. Liquid-phase TEM tracks the local vibrations of NPs around the lattice sites, thereby experimentally measuring the phonon dispersion spectra of different Maxwell lattices following our customized PMN workflow. Integrated with our discrete mechanical model and coarse-grained (CG) inter-NP interaction calculations, these phonon dispersion spectra map the phonon modes and interaction potentials of the lattices, elucidating the relationship between phonon dynamics and colloidal interactions while bringing entropic and many-body effects unique to nanoassemblies to traditional mechanics theories (*25*). Our work delineates the design rules of NP-assembled mechanical frames and demonstrates a collection of unconventional nonlinear lattice deformation paths guided by linear phonon modes for nanoframes, such as gliding, twin boundary migration, and quadruple junction formation, opening new avenues in manufacturing *self-assembled* topologically-engineered mechanical metamaterials. The PMN platform is also applied to hexagonal lattices, whose phonon dispersion spectra are isotropic at long wavelength without floppy modes, suppressing lattice reconfiguration. The platform can be generalized to facilitate phonon engineering at the nanoscale, in systems such as the vastly diverse NP-assembled lattices that have been achieved (*14*), random or amorphous networks (*26*), and nanostructures with controlled defects as next-generation phononic materials.

**Self-assembly of a nanoscale Maxwell lattice with structural degeneracy**

We first choose to study the extreme case of Maxwell lattices, which are at the verge of mechanical instability. In particular, we use shape anisotropy of NPs to control the stability of these Maxwell lattices. Maxwell lattices are characterized by the balance of degrees of freedom of building blocks and constraints from interactions between them (*6, 7*). In the simplified case of point-like building blocks and central-force potentials (radially symmetric), this balance takes the form of $\langle z \rangle = 2d$, where $\langle z \rangle$ denotes the average coordination number of the building blocks and $d$ is the dimension of space (e.g., $\langle z \rangle = 4$ when $d = 2$). In the case of anisotropic NPs, deviations from central-force potentials lead to an interesting contrast between "strong" interactions which only depend on the center-to-center distance between the NPs (usually the nearest-neighboring (NN) bonds in the lattice), and "weak" interactions beyond NNs (e.g., potentials that depend on bond angles). We thus introduce an additional parameter of Maxwellness to measure the ratio of the leading order weak interactions to strong interactions, with twofold importance: First, the mechanical properties of ideal Maxwell lattices are asymptotically approached when this ratio goes to 0, making it a desirable limit to achieve the predicted properties; Second, the stability of the lattices requires weak interactions (without which the lattices have unbounded floppy modes leading to collapsing), which favors a slight non-zero Maxwellness.

We use gold nanocubes with slight corner truncation (Fig. 1b,c and fig. S1) as the building blocks for Maxwell lattices. The gold nanocubes are negatively-charged and dispersed in water due to electrostatic repulsion $E_{\text{el}}$. Liquid-phase TEM imaging is performed at low electron dose rates to minimize beam-induced reactions and to preserve inter-NP interactions (*22*). Increased ionic strength $I$ makes the inter-NP van der Waals attraction $E_{\text{vdw}}$ overwhelm the screened $E_{\text{el}}$, triggering the assembly into a single-layer rhombic lattice of $\langle z \rangle = 4$ (Fig. 1d,e and movie S1). Radial distribution function of the lattice confirms a high crystallinity due to low dispersity of the cubes and self-correction dynamics during assembly (fig. S2 and movie S1). Regarding Maxwell characteristics, the rhombic lattice maintains a stable bond length $l$ and a bimodal distribution of the bond angle θ, suggesting a high enough Maxwellness to ensure structural stability and a low enough Maxwellness to permit floppy modes (Fig. 1b). Following the Boltzmann distribution, the bond angle histogram is converted to a free energy diagram with two potential wells connected by an energy barrier, which can be overcome by thermal fluctuation to allow reconfiguration of rhombuses between degenerate left- and right-leaning states (Fig. 1g and note S2), contrasting from the stable relaxation dynamics of traditional close packings (*22*).

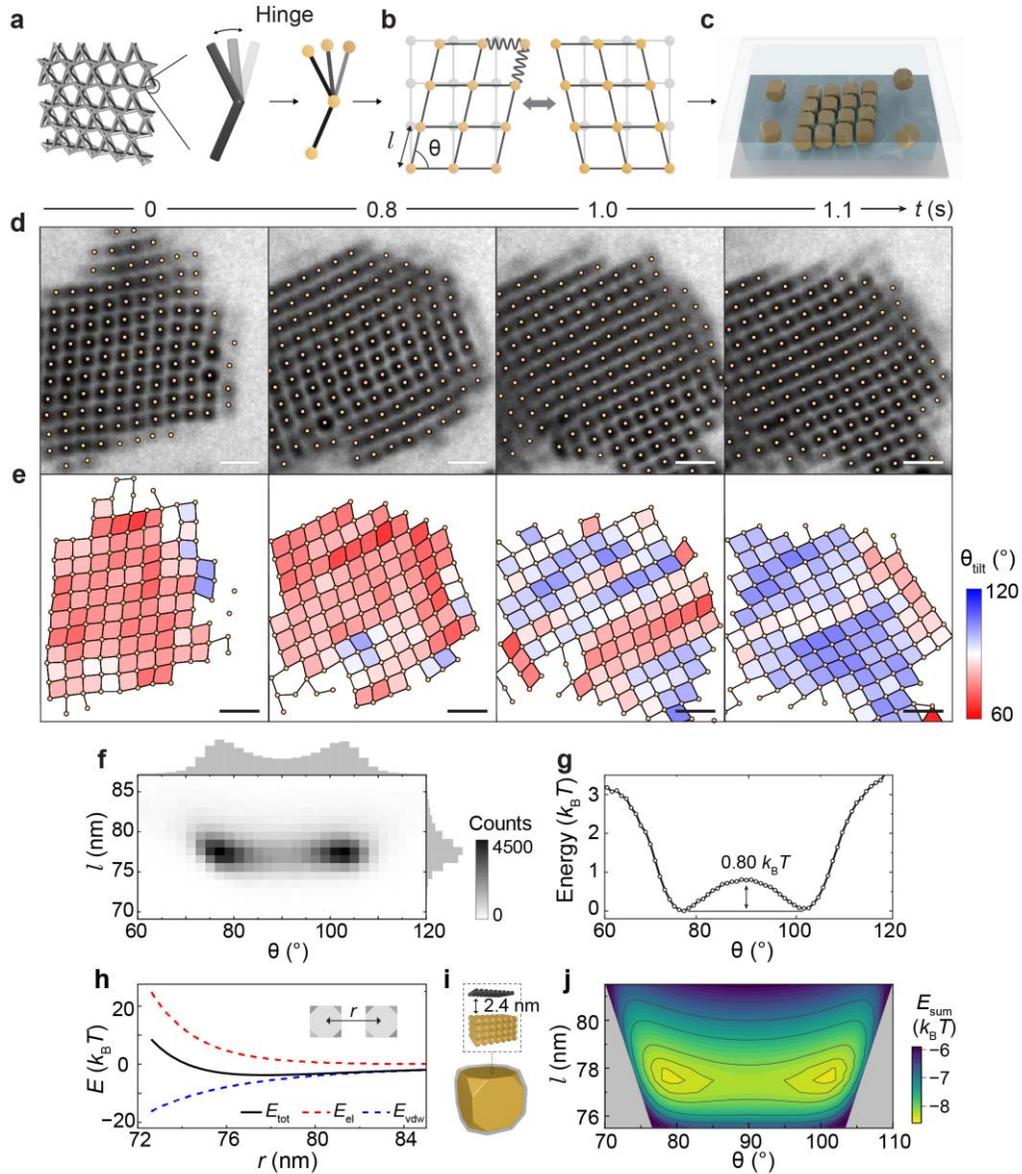

**Fig. 1. Self-assembled Maxwell lattice from gold nanocubes with structural degeneracy. a**, Schematics of macroscopic (kagome bilayer (*28*)) Maxwell lattices and a hinge exhibiting rotational freedom. **b**, Schematic of a rhombic lattice with structural degeneracy. Bond angle θ and bond length $l$ are noted. **c**, Self-assembly of gold nanocubes into a rhombic Maxwell lattice in silicon nitride ($SiN_x$) chambered liquid-phase TEM. **d,e**, Time-lapse liquid-phase TEM images (**d**) overlaid with tracked NP centroids (yellow filled circles) and the bond network with rhombuses colored according to $θ_{tilt}$ (**e**) describing the angle that each rhombus leans towards. Definition of $θ_{tilt}$ in fig. S4. **f**, Histogram of rhombuses of certain $l$ and θ values in movie S1. **g**, Free energy of rhombuses as a function of θ derived from **f** (all $l$ values combined). **h,i**, CG modeling of interaction energy $E_{vdw}$, $E_{el}$ and $E_{tot}$ for a NP pair as a function of the center-to-center distance $r$ (inset schematic) when the side facets of the NPs are aligned (**h**) and schematics of the CG model of a gold nanocube (gold) with ligands (black) (**i**, note S2). **j**, Diagram of total energy $E_{sum}$ of multiple pairs between the central NP and a surrounding NP as a function of θ and $l$ up to the second nearest neighbors, which is sufficient for predicting the stable assembly structure (fig. S5 and note S3). Scale bars: 150 nm.

To understand what inter-NP interactions are important in determining the rhombic lattice self-assembly, we adopt both CG modeling describing the full forms of inter-NP interactions and Brownian dynamics (BD) simulation capturing the assembly process, which pinpoint the role of beyond-NN interactions. Figure 1i shows the CG models of ligand-coated gold cubes (note S2) to calculate the pairwise interaction $E_{\text{tot}}$ between two NPs following $E_{\text{tot}} = E_{\text{vdw}} + E_{\text{el}}$ (Eq. 1) (see an example in Fig. 1h). Given one central NP, while NN-only interactions predict a square lattice ($\theta = 90°$) at the energy minimum, considering next-NNs and beyond converges to energy minima split at two $\theta$ values with a mirror plane at $\theta = 90°$ (Fig. 1j, figs. S5 and S6, notes S2 and S3), matching the structural degeneracy of a rhombic lattice. Zoomed-in view of Fig. 1j reveals an interaction energy landscape (fig. 5c) saddled at a square lattice ($\theta = 90°$), which is metastable and can fall into the global minimum of a rhombic lattice, consistent with our BD simulation of the self-assembly process (fig. S7, note S4, and movie S2). To account for the large size of the system and the complexity of inter-NP interaction (anisotropic and beyond NNs), we develop deep neural network-based machine learning (ML) method for fast prediction of the force and torque exerted on each NP at each time step in our BD simulation (fig. S7 and note S4), which can be generalized to other NP systems. In comparison, previous simulations of cube assemblies considered only NN interactions and predicted square lattices in two dimensions (*29, 30*). The importance of beyond-NN interaction can be a feature of NP systems, where the range of colloidal interactions can be comparable with or even larger than the size of NPs.

**Nanoscale mapping of floppy modes and phonon dispersion spectra by PMN**

In addition to structural transformability, the defining mechanical feature of Maxwell lattices is the existence of floppy phonon modes (*7*), which can take the form of soft planewave modes along certain high-symmetry directions in the bulk phonon spectra (the case of the rhombic lattices that we consider here), or floppy modes exponentially localized at the edges or interfaces of the lattice, depending on the type of the Maxwell lattice (*7*). Inspired by the pioneering work on phonon mode mapping using optical microscopy (*31-33*), we adapt liquid-phase TEM into a PMN platform to directly image phonon dynamics of NP-assembled lattices, where NP positions are resolved at the nanometer resolution.

The fundamental assumption of PMN is that the assembled structure experiences small vibrations around a stable configuration under thermal fluctuation (*33*), which is satisfied in our NP system between structural reconfiguration (Fig. 2a, fig. S8, and note S5). Following equilibrium statistical mechanics, the correlation functions of NP vibrations are related to the dynamical matrix $D$ by $\langle \boldsymbol{u}_i \boldsymbol{u}_{i'} \rangle = k_{\text{B}} T D^{-1}_{i,i'}$ (Eq. 2), where $i, i'$ label the NPs tracked in liquid-phase TEM videos, and $\boldsymbol{u}$ is the displacement vector of a NP per frame from its time-averaged position after drift correction (Fig. 2, b to d, fig. S9, note S6, and movie S3). Phonon structures are then experimentally measured for the first time for NP-assembled lattices from the dynamical matrix in Fourier space $D_{\boldsymbol{k}}$ (a 2×2 matrix, leading to 2 branches of modes), while eigenvectors are derived as polarization of the modes, giving 1 for longitudinal and 0 for transverse waves (Fig. 2f and note S7). The low values of polarization for the lower branch indicate that it is dominated by transverse waves where NPs slide past each other. In contrast, the upper branch is dominated by longitudinal waves, indicating bond length changes and thus higher energies. This PMN method for phonon structures of lattices only requires high spatial resolution tracking of NP vibrations; it does not require theoretical modeling or *a priori* knowledge on the inter-NP interaction as long as the lattice vibrates in equilibrium.

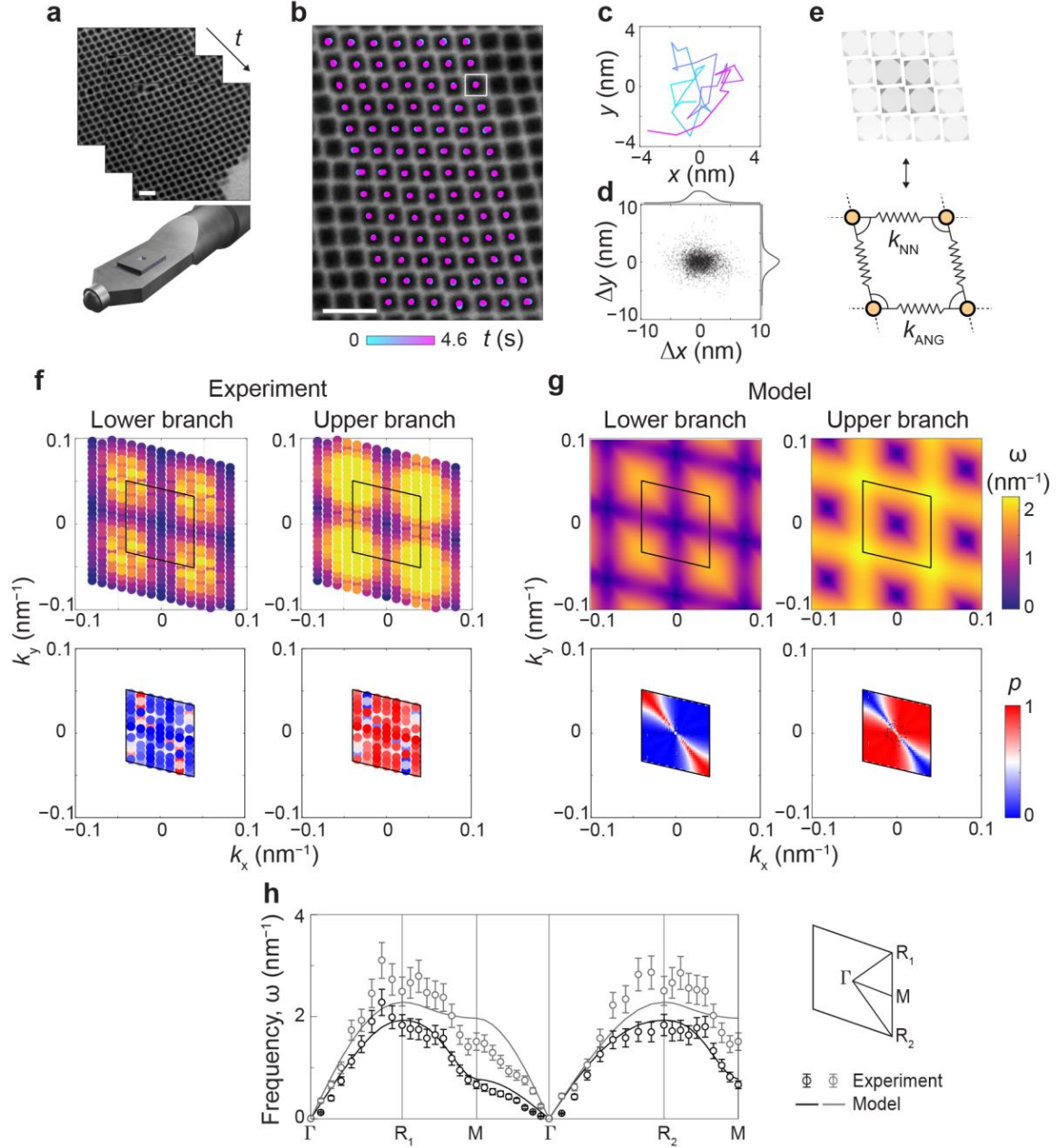

**Fig. 2. Direct imaging of phonon dynamics and integrated theoretical framework to extract phonon dispersion spectra using PMN. a**, Illustration of the time-lapse liquid-phase TEM videos of lattices. **b,c**, Drift corrected trajectory mapped over a selected region (**b**) of a stable rhombus lattice and a zoomed-in single NP trajectory (**c**) corresponding to the NP boxed in white in **b**. The trajectory is colored to elapsed time. **d**, Histogram of the displacements ($\Delta x$ and $\Delta y$) of NPs at a given frame from their time-averaged centroid positions; statistics include all the NPs in **b**. **e**, Discrete mechanical model of one rhombus in the lattice, consisting of four NN springs and four angular springs. **f,g**, Phonon frequencies (top row) and polarizations (bottom row) of experiment (**f**) and model (**g**) plotted based on spring constants fitted following PMN. **h**, Phonon dispersion along high symmetry paths. The lower branch has higher precision since it is a mode of large displacements and is well sampled in liquid-phase TEM experiments. The errors are estimated as detailed in note S7. Results of the lower branch are in black, and the upper branch in gray. Scale bars: 200 nm.

**Phonon behaviors of NP-assembled lattices determined quantitatively by complex inter-NP interactions**

To understand the experimentally measured phonon dispersion spectra, we develop a discrete mechanical model (Fig. 2e, note S7, and table S2) to describe the NP-assembled lattices as mass beads connected by "effective" springs. This model consists of two sets of potentials, one connecting NNs representing the strong interaction, with $k_{NN}$ as the harmonic spring constant, and the other a bistable angular (ANG) potential to account for the weak interaction which are beyond-NN interactions, three-body, and anharmonic, with $k_{ANG}$ as the strength of the ANG potential. Here $k_{ANG}$ controls both the stiffness of the ANG spring when expanded around the local minima and the barrier between them, an approximation justified by expanding the potential locally around θ = 90°. Compared to traditional models of Maxwell lattices, we account for the complexity of inter-NP interactions by not only introducing the three-body ANG potentials but also capturing the intrinsic bistability of this potential: The two minima of $V_{ANG}(θ)$ at $θ_0$ and $(180° - θ_0)$ correspond to the two degenerate, left- and right-leaning rhombus configurations.

This step of PMN fits the dynamic matrix calculated from experimental tracking of NPs to that predicted from our discrete mechanical model, to extract spring stiffness and Maxwellness of the lattice. For clarity, a phonon dispersion is plotted along the high symmetry paths in this example (Fig. 2h). A branch of floppy modes with low frequency sits between the Γ and M points in the first Brillouin zone, corresponding to the zero modes in an ideal rhombic lattice and serving as a clear signature that the lattice is close to an ideal Maxwell lattice. The highest measurable phonon frequency by PMN is on the order of GHz determined by the tracking precision of NPs, varied upon the mass of NPs (fig. S10 and note S1). The fitting (Fig. 2f,g and note S7) yields $k_{NN} = (0.971 \pm 0.012)\ k_B T\ \text{nm}^{-2}$ and $k_{ANG} = (4.99 \pm 0.15) \times 10^{-5}\ k_B T\ \text{deg}^{-4}$. Physically, these parameters describe that a mechanical deformation associated with an energy of 1 $k_B T$ leads to a longitudinal stretching of a NN bond as small as 1.4 nm and an angular fluctuation as high as 8.3°. Maxwellness $M$ is calculated by $M = k'_{ANG}/k_{NN}$ (Eq. 3), where $k'_{ANG}$ is normalized $k_{ANG}$ to make $M$ dimensionless (fig. S10, note S7, and table S2). In this example, $M = 0.0383 \pm 0.0012$, suggesting that the system is close to an ideal Maxwell lattice yet robust under thermal fluctuation. As a comparison, for macroscopic structures, $M = 0$ is often achieved with high bulk modulus compared with hinge friction (*9*). Our result of a NP-assembled Maxwell lattice demonstrates a promising design space of materials with both longitudinal stability and rotational freedom at the hinges. Rotational degrees of freedom of the NPs are excluded from this fitting by representing the NPs as point-like particles in the model. The NP rotations are observed to be tightly confined and provide small entropic corrections to the spring constants (figs. S11 and S12, note S8, and table S3).

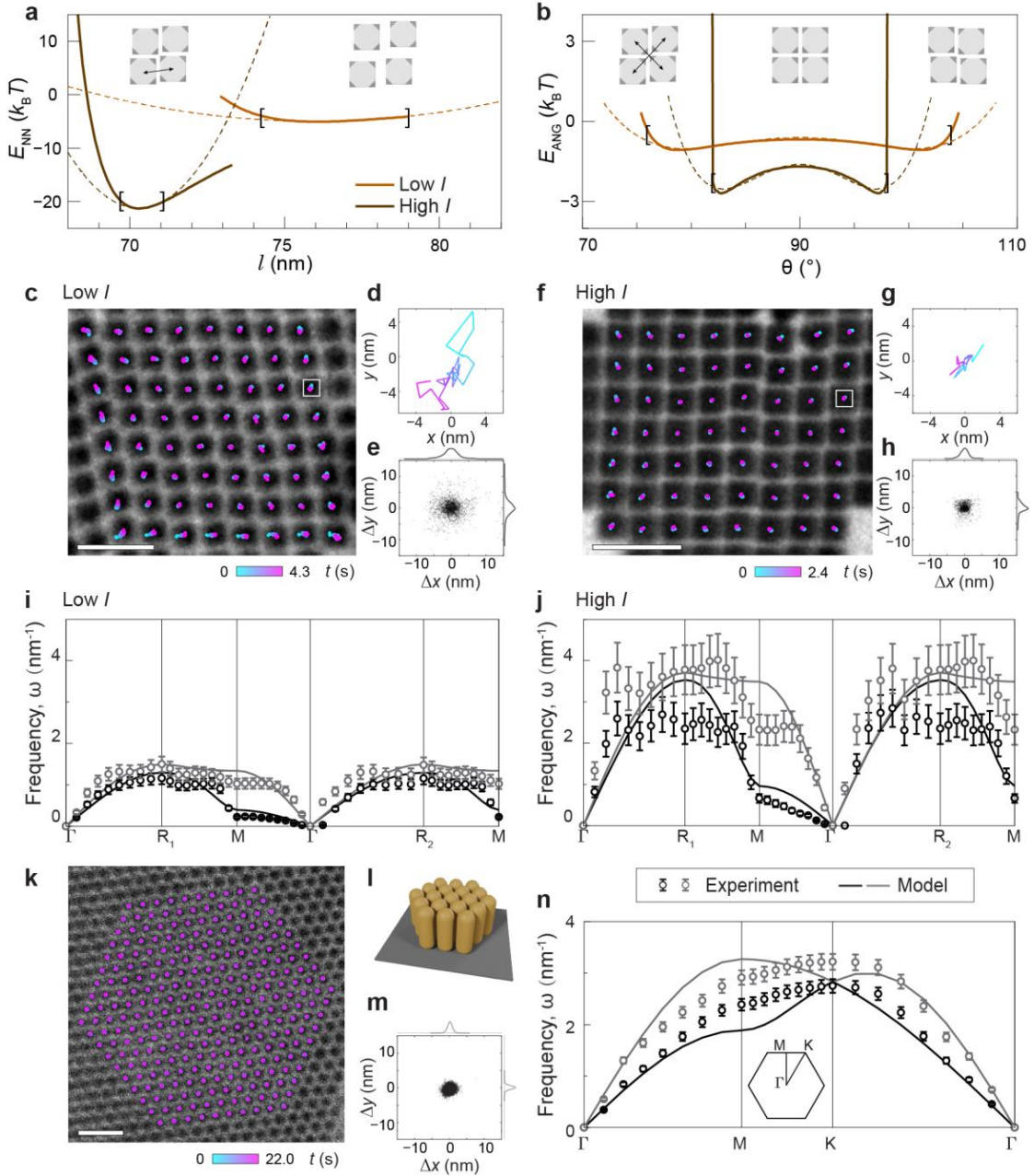

**Fig. 3. Effects of inter-NP interactions on NP vibrations, phonon modes, and spring constants. a,b**, CG-modeled interaction energy of NN springs as a function of $l$ (**a**) and interaction energy of ANG springs as a function of $\theta$ (**b**). The dashed lines are fittings to the potential forms of the discrete mechanical models (Eqs. 3,4). The brackets indicate the $[E_{NN,m}, E_{NN,m} + 1\ k_BT]$ fitting range in **a** and $[E_{ANG,m}, E_{ANG,m} + 1\ k_BT]$ in **b**, where $E_{NN,m}$ is the minimum of the sum of interaction energy of all four NN pairs $E_{NN}$, and $E_{ANG,m}$ is the minimum of the sum of interaction energy of two diagonal pairs $E_{ANG}$ (insets, note S9). **c–h**, Liquid-phase TEM image overlaid with drift corrected trajectories colored to elapsed time (**c,f**), a representative trajectory of one particle (**d,g**) and displacement distribution of drift corrected centroids of all NPs (**e,h**) in a Maxwell lattice at low (**c–e**) and high ionic strength (**f–h**). **i,j**, Phonon dispersion spectra along high symmetry paths of low (**i**) and high ionic strength (**j**). **k,l**, Liquid-phase TEM image (**k**) and schematic (**l**) of standing gold nanorods assembling into a hexagonal lattice overlaid with drift corrected

trajectories colored to elapsed time. **m**, Displacement distribution of the drift corrected centroids of all nanorods. **n**, Phonon dispersion spectra along high symmetry paths for nanorod assembly. Scale bars: 200 nm.

It is noteworthy that the experimentally mapped phonon modes via PMN fit well with our discrete mechanical model at the harmonic level, although inter-NP interactions can exhibit far more complexity than harmonic potentials and are sensitive to details such as NP surface chemistry and shapes. To rationalize the agreement, we study comparatively the sets of $k_{\mathrm{NN}}$ and $k'_{\mathrm{ANG}}$ derived from the CG-modeled inter-NP interaction and from our liquid-phase TEM experiments (using the PMN workflow) for rhombic lattices assembled at different ionic strengths (low and high, fig. S13 and movie S4). In CG models that construct inter-NP interactions in an *ab initio* manner, within 1 $k_\mathrm{B}T$ of fluctuations around the equilibrium lattice site, the NN interaction is well fitted by a harmonic potential with spring constant $k_{\mathrm{NN}}$ and the ANG potential is well fitted by a quartic double-well form of coefficient $k_{\mathrm{ANG}}$, supporting the discrete mechanics model (Fig. 3a,b, note S9, and table S4). The fitted values of $k_{\mathrm{NN}}$ and $k'_{\mathrm{ANG}}$ from the CG model and from the liquid-phase TEM imaging agree qualitatively (Fig. 3, a to j, and Table 1), which suggests that thermal fluctuations smoothen the complex forms of inter-NP interactions at the nanoscale, permitting simple effective models for lattice dynamics. The slight quantitative discrepancies in the $k_{\mathrm{NN}}$ and $k'_{\mathrm{ANG}}$ values from CG model and liquid-phase TEM imaging come from limitations both in modeling complex inter-NP interactions and statistical sampling in PMN. As $I$ increases, $k_{\mathrm{NN}}$ and $k'_{\mathrm{ANG}}$ significantly increase, indicating steeper inter-NP interaction potentials. The variations of inter-NP interactions manifest as the coefficients controlling the mechanical strength and flexibility of nanoframes.

| $I$ (mM) | $k_{\mathrm{NN}}$ ($k_\mathrm{B}T$/nm²) | | $k'_{\mathrm{ANG}}$ ($k_\mathrm{B}T$/nm²) | | $M$ | |
|---|---|---|---|---|---|---|
| | PMN | CG | PMN | CG | PMN | CG |
| Low | 0.450 ±0.007 | 0.219 ±0.017 | 0.0097 ±0.0005 | 0.0044 ±0.0002 | 0.0216 ±0.0012 | 0.0201 ±0.0018 |
| High | 3.05 ±0.06 | 4.18 ±0.47 | 0.0575 ±0.0045 | 0.0259 ±0.0008 | 0.0189 ±0.0015 | 0.0062 ±0.0007 |

**Table 1**. **Comparison of spring stiffness and Maxwellness derived from PMN and CG model at low and high ionic strength.**

The PMN workflow is generally applicable to other NP self-assemblies. In our study of a hexagonal lattice self-assembled from gold nanorods (Fig. 3, k to m) as an example of close-packed non-Maxwell lattices ($\langle z \rangle = 6 > 2d$, "over-constrained"), no floppy modes should arise in its spectra except the trivial translational modes at $k = 0$, and the phonon structure should be isotropic at small $\boldsymbol{k}$. The measured phonon spectra in Fig. 3n agrees well with these expectations and is well-fitted by a simple mechanical model with harmonic springs between NNs.

**Nonlinear lattice reconfiguration upon agitation characteristic of a Maxwell lattice**

The floppy modes in Maxwell lattices are easily excited to large amplitudes, leading to collective nonlinear lattice deformation paths upon agitation. As shown in Fig. 4a,f, the low kinetic barrier for a rhombus to switch between left- and right-leaning states given the anharmonic potential (Fig.

1g) underpins the elementary motion of "layer gliding" where one row (or sub-row) of cubes glide concertedly. Gliding creates a contact twin, which can be dynamically driven by thermal fluctuations at the twin boundary (TB). Subsequent shifting of the orientations of other rhombus rows pushes the migration of the TB (Fig. 4a and movie S5). Such spatially localized TB shifting starts from a linear combination of the floppy modes along the Γ and M points in the first Brillouin zone (Fig. 4, b to e) and continues to the nonlinear regime. Although twinning and the associated dynamics have been found in crystalline minerals or metals (*34*), such as deformation twinning upon an external stress (*35*), this is the first-time observation of twinning dynamics in NP-assembled lattices. When the displacement of gliding is multiples of bond length, no twinning is created; instead, the lattice symmetry is maintained across the gliding plane (fig. S14, c to e).

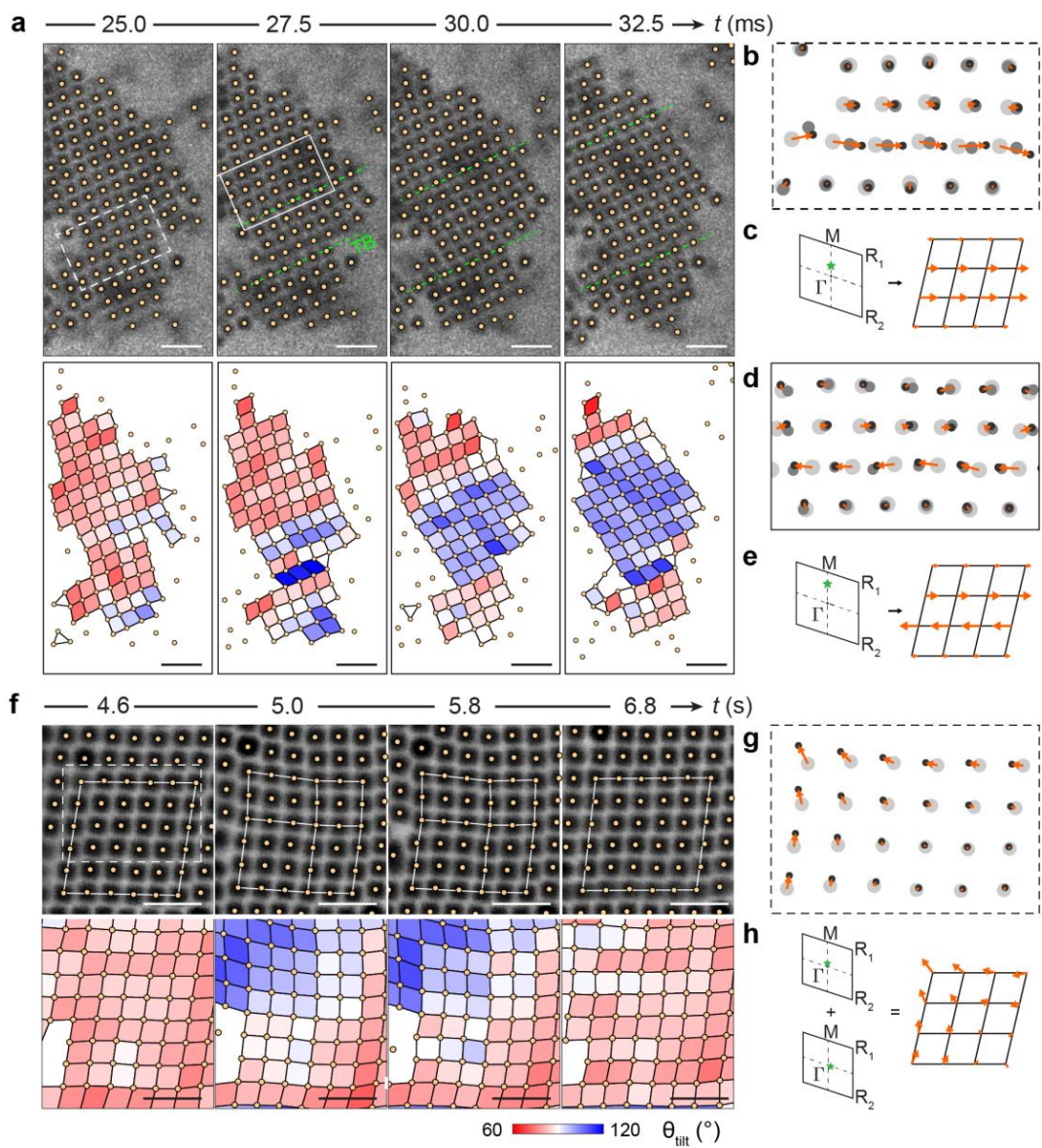

**Fig. 4. Collective deformation paths of Maxwell lattices upon thermal agitation. a,f**, Time-lapse liquid-phase TEM images overlaid with tracked NP centroid positions (top), and rhombuses colored by $\theta_{tilt}$ (bottom). The TB (**a**, labelled as green dashed lines) migration process is captured by direct electron

detector at a frame rate of 400 frames per second (fps). The quadruple junction formation and annealing process (**f**) is captured with Orius camera. **b,d**, Zoomed-in view of NP motions in the dashed (**b**) and solid boxes (**d**) in **a**, highlighting gliding. **c,e**, Schematics of the first Brillouin zone and one of the phonon modes corresponding to the positions marked by the green stars. **f,g**, Zoomed-in view of NP motions in the dashed box in **f**, highlighting quadruple junction formation. **h**, Schematic of the first Brillouin zone and superposition of the phonon modes corresponding to the positions marked by the green stars. Scale bars: 200 nm.

When multiple layers (either parallel or intercepting) glide together, large scale collective twinning (fig. S14a,b) or quadruple junction formation are observed. Figure 4f shows the formation and annealing of a quadruple junction where twins intercept and pin. This leads to polysynthetic twinning (*36*), characterized by a plurality of contact twins with parallel or nonparallel layers. Previous simulation studies have suggested this relaxation mode in DNA-coated colloidal lattice (*37*), which has not been verified experimentally. This lattice relaxation to linear order is a combination of phonon modes along the valleys in the phonon spectra in Fig. 2f,g, corresponding to transverse motions (Fig. 4g,h), contrasting with the stable relaxation dynamics limited to lattice surface or local defect observed in previous studies (*22*). Conventionally the effect of TB on materials' mechanical strength is understood in metal systems in two ways (*38*). The generation and migration of TB can effectively soften the material, while TB can also strengthen the material by acting as pinning points impeding dislocation propagation, known as the Hall-Petch effect. Our observation of quadruple junction might suggest that the TB strengthening mechanism does not hold for the nanoscale Maxwell lattice.

**Conclusion**

We bridge topologically-engineered mechanical frames with NP self-assembly, all the way from experimental realization of Maxwell nanoframes to direct spatially-resolved phonon imaging and establishment of a theoretical framework predicting phonon dynamics from fundamental inter-NP interactions. Our Maxwell lattices suspended in liquids are overdamped, exhibiting structural reconfigurability and a frequency range of 0–117 MHz (~ultrasound), with floppy mode features at 20 MHz. Phononic devices at this frequency range are of interest for applications in noninvasive imaging, sensing, acoustic waveguide, and diode for sonar cloaking. When the same assemblies are dry, we expect the modes to become underdamped, and appear at higher frequencies (about GHz scale) due to the lack of screening of inter-NP interactions.

Here the floppy modes are unambiguously captured by our PMN platform. At the nonlinear level, the floppy modes lead to various lattice deformation paths captured for the first time upon external perturbation, which are modulated by inter-NP interactions (e.g., ionic strengths for charged NPs; figs. S15 and S16, note S10). Technically, the PMN platform based on liquid-phase TEM can extend to other NP-based assemblies, including not only lattices of diverse symmetries but amorphous glass or networks for quantitative phonon manipulation. We foresee enormous opportunities in "*self-assembled mechanical metamaterials*" given (i) the ever-increasing libraries of topologies that can be achieved by diverse self-assembly strategies such as solvent-evaporation, binary mixing and out-of-equilibrium pattern formation, (ii) new theories to be established that can account for more complexity of nanoassemblies (e.g., chirality, mixtures, amorphous structures) and reconfigurability, (iii) the essentially limitless selections of NP building blocks of different size, composition, surface chemistry as well as plasmonic, electronic, catalytic, and magnetic properties for emergent functions beyond mere topology-encoded mechanical properties.

**Acknowledgements**

**Funding:** Experiments and theoretical modeling for this work were supported by the Office of Naval Research (MURI N00014-20-1-2479). The BD simulation effort of this work was supported by the Defense Established Program to Stimulate Competitive Research (DEPSCoR) Grant No. FA9550-20-1-0072.

**Author information**: These authors contributed equally: Chang Qian, Ethan Stanifer

**Author Contributions**: C.Q. and Q.C. designed the experiments. C.Q., and B.L. performed the experiments. C.Q., L.Y., C.L. and Q.C. did the CG modeling and single particle tracking analysis. E.S. and X.M. developed the discrete mechanical model and theory. C.Q., E.S. and X.M. performed PMN analysis. Z.M. and W.P. performed the BD simulations with ML-based inter-NP interaction modeling. All authors contributed to the writing of the paper. Q.C., X.M. and W.P. supervised the work.

**Competing interests:** The authors declare no competing interests.

**Correspondence and requests for materials** should be addressed to Xiaoming Mao, Qian Chen.